\documentstyle{grf1996b}

\date{March 2004}

\begin{document} 

\bigskip

\begin{frontmatter}

\title{Neutrino oscillations and supernovae}

\author{D. V. Ahluwalia-Khalilova}
\address{ISGBG, Department of Mathematics, Ap. Postal C-600\\ 
University of Zacatecas (UAZ), Zacatecas 98060, Mexico} 

\begin{abstract}
In a 1996 JRO Fellowship Research Proposal (Los Alamos),
the author suggested that neutrino oscillations
may provide a powerful indirect energy transport mechanism 
to supernovae explosions.  The principal aim of this
addendum is to present the 
relevant \textit{unedited} text of Section 1
of that proposal.
We then briefly remind, (a) of an early suggestion of
Mazurek on vacuum neutrino oscillations 
and their relevance to supernovae explosion, and (b)
Wolfenstein's result on suppression of the effect by
matter effects. We conclude that
whether or not neutrino oscillations play a significant role
in supernovae explosions shall depend if there are shells/regions 
of space 
in stellar collapse where matter effects play no 
essential role.
Should such regions exist in actual astrophysical situations,
the final outcome of neutrino oscillations on 
supernovae explosions shall depend, in part, on whether or not the 
LNSD signal is confirmed.
Importantly, the reader is reminded that neutrino oscillations
form a set of flavor-oscillation clocks and these clock 
suffer gravitational redshift which can be as large as 
20 percent. This effect must be incorporated fully into
any calculation of supernova explosion.
\end{abstract}

\end{frontmatter}

\section{Section 1 of author's 1996 
 J Robert Oppenheimer  Fellowship Research Proposal: 
Unedited text}

The following is an unedited text of Section 1 of author's 1996 
J Robert Oppenheimer  Fellowship Research Proposal:\footnote{
The reader is reminded that such proposals are written for a very broad 
readership, as the evaluators come not only from physics, but fields
as far as biology, and other sciences.       
The proposal was submitted 
by M. B. Johnson in July 1996 in
his ``Nomination of Dharam Ahluwalia for Oppenheimer 
Fellowship.'' The addendum title coincides with the 
title of Section 1 of the proposal.}

Neutrinos were introduced in physics by Pauli to save conservation 
of energy and momentum in the $\beta$-decay: Neutron $\rightarrow$ 
Proton + Electron + Anti-electron Neutrino. All the planets and  galaxies are embedded in a sea of neutrinos with a number density of roughly 100 neutrinos/cm$^3$. Our own Sun shines via thermonuclear processes 
that emit neutrinos in enormous number. Because of their weak interactions, neutrinos, unlike photons, can pass through extremely dense matter very efficiently. This fact makes neutrinos primary agents for energy transport 
in the dense matter associated with supernovae and neutron stars.

  Since their initial experimental observation by Frederick Reines 
and C. L. Cowan, neutrinos are now known to exist in three types. 
These types are called ``electron,'' ``muon,'' and ``tau'' and 
are generically written as $\nu_e$, $\nu_\mu$, and $\nu_\tau$. 
A series of empirical anomalies indicates that  neutrinos may not have a definite mass but, instead, be in a linear superposition of three different mass eigenstates. The mass differences in the underlying mass eigenstates would cause a neutrino of one type to ``oscillate'' to a neutrino of another type as may have been seen recently at the LSND neutrino oscillation experiment at LANL. The phenomenon of neutrino oscillations, if experimentally confirmed, will have profound consequences not only for nuclear and particle physics but also for astrophysics and cosmology.

   I have already noted the neutrinos to be prime drivers of supernova explosions. The phenomenon of neutrino oscillations will alter the evolution of supernova explosion. The basic problem that still stands unsolved is a robust theory of supernova explosions. In the context of supernova explosions, and the problem of obtaining successful explosions, I now follow Colgate 
\textit{et al.} 
[S. A. Colgate, M. Herant, and W. Benz, Phys. Rep. {\bf 227}, 157 (1993)] 
and assume that the matter next to the neutron star is heated by neutrinos from the cooling neutron star. They note that in some models ``this result in strong, large scale convective flows in the gravitational field of the neutron star that can drive successful, albeit weak, explosions.'' I emphasize that all authors find that without ``fine tuning'' the explosions are weak and lack about five percent of the energy needed for an explosion. Qualitatively, this missing energy needed for a robust model of explosion may be provided if the length scales over which neutrino oscillations take place are of the same order of magnitude as the spatial extent of a neutron star and neutrino-sphere, because while

\begin{quote}
        the energy flux in each of the electron neutrinos and antineutrinos is 
        about $L_{\nu_e}\approx   L_{\overline{\nu}_e}\approx$   
few$\times$  $10^{52}$   ergs s$^{-1}$, 
with comparable fluxes of $\nu_\mu$, $\overline{\nu}_\mu$,
$\nu_\tau$, and  $\overline{\nu}_\tau$,
\end{quote}
the 
\begin{quote}
        average energy of $\nu_e$  
is about 10 MeV, the average energy of other 
        neutrinos is \textit{higher} 
by a factor of 2 for $\nu_\mu$    and   $\overline{\nu}_\mu$, 
and by a factor of 3 for    $\nu_\tau$, and  $\overline{\nu}_\tau$.
\end{quote}
Any oscillation between neutrinos of different flavors is, therefore, an indirect energy transport mechanism towards the actively interacting
$\nu_e$    and   $\overline{\nu}_e$.
Qualitatively this contributes in the direction of the robustness of the explosion. My resent work, with C. Burgard, on the solution of terrestrial neutrino anomalies provides precisely the neutrino oscillation 
parameters that yield the oscillation length scales of just the right order of magnitude for supernova physics (and in addition predict the observed solar neutrino deficit).

   In order to make these qualitative arguments quantitative two 
additional physical 
processes affecting the above indicated \textit{vacuum}
neutrino oscillations must be incorporated: (a) The presence of large electron densities in astrophysical environment makes it necessary that relevant matter induced effects, suggested by Mikheyev, Smirnov and 
Wolfenstein, be considered, 
and (b) My work, with C. Burgard, on gravitationally induced neutrino oscillation phases also indicates  that strong gravitational fields associated with neutron stars may introduce important modifications to neutrino oscillations, and hence to the suggested energy transport mechanism via neutrino oscillations. As part of my JRO studies I propose to implement the above outlined program quantitatively. My quantitative and qualitative studies so far give reasons to claim that there is every 
 physical reason to believe that the ``missing energy'' in the non-robust models for supernova explosion, the anomaly in the observed deficit in the solar neutrino flux, the excess 
$\overline{\nu}_e$
  events seen at 
LSND at Los Alamos, and the anomaly associated with atmospheric neutrinos, 
\textit{all} 
arise from the same underlying new physics \textemdash 
the phenomenon of neutrino oscillations from one type to another. It is of profound physical importance to place these suspected physical connections on firm quantitative foundations.

\section{Brief remarks}

The above 1996 proposal was a logical continuation, and directly 
connected to, a work jointly done the same year with C. Burgard. 
It has been widely known informally, without a full
access to its text. This addendum fills the gap 
of its availability. 

While writing these remarks the author
has learned that the effect of neutrino oscillations 
on supernovae explosions was first presented in a talk by Mazurek 
\cite{Mazurek:tt}. Soon afterward, Wolfenstein 
 showed that for 
collapsing stellar cores, matter effects dramatically suppress
neutrino oscillations \cite{Wolfenstein:1979ni}, 
with the following one-line \textit{abstract},
``It is shown that even if neutrino oscillations exist
they are effectively inhibited from occurring in collapsing
stars because of the high matter density.''

The suggestion that neutrino oscillations may play a significant
role in supernovae explosions has been pursued vigorously, 
though often without
acknowledging the proposal  of Sec. 1, or
talk of Ref. \cite{Mazurek:tt}.

The matter is far from 
settled, see, e.g, 
\cite{Goswami:bv,Hix:2003vv,Liebendoerfer:2003es,Ahriche:2003wt,Fryer:2003}
and whether or not neutrino oscillations play a significant role
in supernovae explosions shall depend if there are shells/regions 
of space in stellar collapse where matter effects
\cite{Mikheyev:dy,Wolfenstein:1977ue} play 
no essential role. The spatial extent of these 
regions, we suspect, must 
be significantly smaller than neutron-star size.  
Should such regions exist in actual astrophysical situations,
the final outcome of neutrino oscillations on 
supernovae explosions shall depend, in part, whether or not the 
LNSD signal is confirmed \cite{Athanassopoulos:1997pv}. This is so
because the LSND-suggested $\Delta m^2$ alone gives
\textit{smallest}, by about three orders of magnitude
(as compared with the solar and atmospheric data 
\cite{Davis:2002npa,Koshiba:2002npb,Toshito:2001dk}),
required vacuum-oscillation length. 
Assuming a $\Delta m^2 \approx 0.4\;\mbox{eV}^2$ \textendash~
as discussed 
on numerous occasions as early as 1994 at Los Alamos by 
one of us,\footnote{It was recently confirmed in Ref. \cite{Latimer:2003hq}.
The ``$\Delta m^2_{12} \leftrightarrow \Delta m^2_{23}$'' symmetry
noted in  \cite{Latimer:2003hq}, though important, is not new. It
was already proved analytically 
in Ref. \cite{Ahluwalia:1996fy}.}
and noted in Ref. \cite{Ahluwalia:1996ev} \textendash~ 
the obtained length ranges from about 10 meters for
a 1 MeV neutrino, to  300 meters for a 25 MeV neutrino.              

Importantly, it is to be noted that neutrino oscillations
form a set of flavor-oscillation clocks and these clock 
suffer gravitational redshift which can be in the neighborhood 
20 percent. This effect must be incorporated fully into
any calculation of supernova explosion.
References \cite{Konno:1998,Wudka:2000rf,Crocker:2003cw,Linet:2002wp,Adak:2000tp,Grossman:1996vp} shed further light on 
Ref. \cite{Ahluwalia:1996ev} and deal with impact of gravity 
on neutrino oscillations.

\end{document}